\documentclass{article}

\usepackage{PRIMEarxiv}

\usepackage[utf8]{inputenc} 
\usepackage[T1]{fontenc}    
\usepackage{hyperref}       
\usepackage{url}            
\usepackage{booktabs}       
\usepackage{amsfonts}       
\usepackage{nicefrac}       
\usepackage{microtype}      
\usepackage{lipsum}
\usepackage{fancyhdr}       
\usepackage{graphicx}       
\graphicspath{{media/}}     
\usepackage{amsmath,amssymb,amsfonts,amsthm}
\usepackage{algorithmic}
\usepackage{textcomp}
\usepackage{xcolor}
\usepackage{blindtext}
\usepackage{subfig}
\usepackage{cuted}
\usepackage{physics}
\usepackage{xcolor}
\usepackage[long]{optidef}
\usepackage[detect-none]{siunitx}
\usepackage{multirow}
\usepackage{array}

\usepackage{makecell}
\usepackage{cite}
\usepackage{amsmath,amssymb,amsfonts}
\usepackage{algorithmic}
\usepackage{graphicx}
\usepackage{textcomp}
\usepackage{xcolor}
\DeclareMathOperator{\sinc}{sinc}
\def\BibTeX{{\rm B\kern-.05em{\sc i\kern-.025em b}\kern-.08em
    T\kern-.1667em\lower.7ex\hbox{E}\kern-.125emX}}
\title{Wavelet Packet Division Multiplexing (WPDM)-Aided Industrial WSNs
\thanks{\textit{This material is based upon work supported by Science Foundation Ireland (SFI) and is co-funded under the European Regional Development Fund under Grant Numbers 13/RC/2077 and 13/RC/2077-P2. \\
\copyright 2023 IEEE. Personal use of this material is permitted. Permission from IEEE must be obtained for all other uses, in any current or future media, including reprinting/republishing this material for advertising or promotional purposes, creating new collective works, for resale or redistribution to servers or lists, or reuse of any copyrighted component of this work in other works.}} 
}

\author{\Large{\emph{Indrakshi Dey and Nicola Marchetti}}}

\begin{document}
\maketitle

\begin{abstract}
Industrial Internet-of-Things (IIoT) involve multiple groups of sensors, each group sending its observations on a particular phenomenon to a central computing platform over a multiple access channel (MAC). The central platform incorporates a decision fusion center (DFC) that arrives at global decisions regarding each set of phenomena by combining the received local sensor decisions. Owing to the diverse nature of the sensors and heterogeneous nature of the information they report, it becomes extremely challenging for the DFC to denoise the signals and arrive at multiple reliable global decisions regarding multiple phenomena. The industrial environment represents a specific indoor scenario devoid of windows and filled with different noisy electrical and measuring units. In that case, the MAC is modelled as a large-scale shadowed and slowly-faded  channel corrupted with a combination of Gaussian and impulsive noise. The primary contribution of this paper is to propose a flexible, robust and highly noise-resilient multi-signal transmission framework based on Wavelet packet division multiplexing (WPDM). The local sensor observations from each group of sensors are waveform coded onto wavelet packet basis functions before reporting them over the MAC. We assume a multi-antenna DFC where the waveform-coded sensor observations can be separated by a bank of linear filters or a correlator receiver, owing to the orthogonality of the received waveforms. At the DFC we formulate and compare fusion rules for fusing received multiple sensor decisions, to arrive at reliable conclusions regarding multiple phenomena. Simulation results show that WPDM-aided wireless sensor network (WSN) for IIoT environments offer higher immunity to noise by more than 10 times over performance without WPDM in terms of probability of false detection.
\end{abstract}

\section{Introduction}
Industrial Internet-of-Things (IoT)/Industry 4.0 conceptually will interconnect everything within an industry, including employees, machines and products. They will interact with each other to ensure seamless production without any human intervention. The idea is to deploy different kinds of sensors and related devices that will collect information regarding customer requirements, conditions of the operating machines, and the surrounding conditions in which they operate. Through groups of heterogeneous sensors, machines will communicate among themselves through a virtual social network and will communicate with human beings (managers, owners, customers) through an actual digital network \cite{1,2,3}.

Deploying a large-scale wireless sensor network (WSN) that can enable the deployment of industrial IoT is extremely challenging owing to different factors. All sensors have to transmit their decisions simultaneously over a multiple access channel (MAC) to a decision fusion center (DFC) owing to limited bandwidth and as a result, the DFC receives a superposition of heterogeneous sensor decisions \cite{4}. Such superposition results in interfering sensor signals corrupted with noise. Besides, the wireless MAC suffers from random time-varying fading and shadowing \cite{5}. Additionally, industrial environment is different from traditional indoor environments like homes, offices and buildings. Owing to the large dimensions and presence of noisy instruments and machines, industrial environments suffer from impulsive noise in addition to background and electronic Gaussian noise \cite{6}.

In order to combat the above-mentioned challenges concomitant with large-scale industrial sensing networks, we propose a robust wireless sensor communication system that combines wavelet packet division multiplexing (WPDM) on the transmit side with distributed multi-antenna based decision fusion (DF) on the receive side. Implementing WPDM has been recommended in \cite{7,8} for providing substantial immunity to both impulsive and Gaussian noise. Furthermore, deploying multiple antennas at the DFC can improve fusion performance over deep faded and shadowed MAC \cite{9,10}. Therefore, combining WPDM and multi-antenna DF can provide considerable reduction in vulnerability to noise, fading and shadowing in a large-scale industrial sensing network. 

The primary contribution of this paper is to propose i) waveform coding of local sensor observations from each group of sensors onto wavelet packet basis functions before transmitting them over wireless MAC and ii) fusing the decisions at multi-antenna DFC using linear filter-bank or correlator receiver for separating the received waveform-coded sensor observations, with the aim of incorporating resilience to noise, fading and shadowing. Another major contribution of the paper is to evaluate the performance of our proposed system against a realistic industrial environment, which suffers from a combination of Gaussian distributed additive background noise and impulsive noise. We formulate impulsive noise as either Middleton Class A distributed or Bernoulli-Gaussian distributed \cite{11,12}. Local sensor decisions from each sensor group within an industrial sensing environment are coded with wavelet packet basis functions, and each group is waveform-coded at each level of binary tree structure thereby formulating a WPDM system. The WPDM multiplexed signals are transmitted over the MAC to a DFC equipped with multiple antennas. At the DFC, we formulate linear multi-rate filter-based optimum and sub-optimum fusion rules generalized to the WPDM set-up. Performance of our formulated system is simulated and compared for three different wavelet scaling functions \cite{13}, Haar, Shannon and Piecewise linear spline, in presence of varying degrees of impulse noise. 

The rest of the paper is organized as follows. Section II introduces the system design, Section III formulates the filters and the fusion rules, Section IV analyzes performance of our proposed system, while concluding remarks are provided in Section V.

\section{System Design}

The infographic representation for a large-scale industrial sensing network consisting of heterogeneous sensor groups is provided in Fig.~\ref{fig2a}. We assume that there are different groups of sensors; $z \triangleq \{1, \dotso, Z\}$. For example, one group of sensors could be mobility sensors reporting if there is a moving target or not, another group could sense if a certain harmful chemical is present or not, another group could sense if carcinogenic material like asbestos is present or not etc. Each of the $Z$ sensor groups transmit their observations to a DFC equipped with $N$ antennas. The observations are waveform coded onto wavelet packet basis functions, thereby implementing a wavelet packet division multiplexing (WPDM) structure. 

\subsection{System Model}

We design a binary tree structure with $L$ number of levels such that $Z = 2^L$ or $L = \log_2 Z$. If $Z$ is odd, then $L = \lceil \log_2 Z \rceil$, where $\lceil \rceil$ denotes the ceiling function. Let us also assume that the $z$th sensor group consists of $M$ transmit sensors $(m \triangleq \{1, \dotso, M\})$. Each of the $M$ sensors in each group transmits its independent observation over a $T_l$ symbol duration using a time division multiplexed system, where $l$ is the level of the binary tree structure of the WPDM system, $l \geq 0; l \triangleq \{1, \dotso, L\}$. Here $T_l = 2^l T_0$ where $T_0$ is the time interval between symbols. It is worth-mentioning here that the tree can be pruned and developed depending on the total number of sensor groups present. 
\begin{figure}[t]
\begin{center}
 \includegraphics[width=0.7\linewidth]{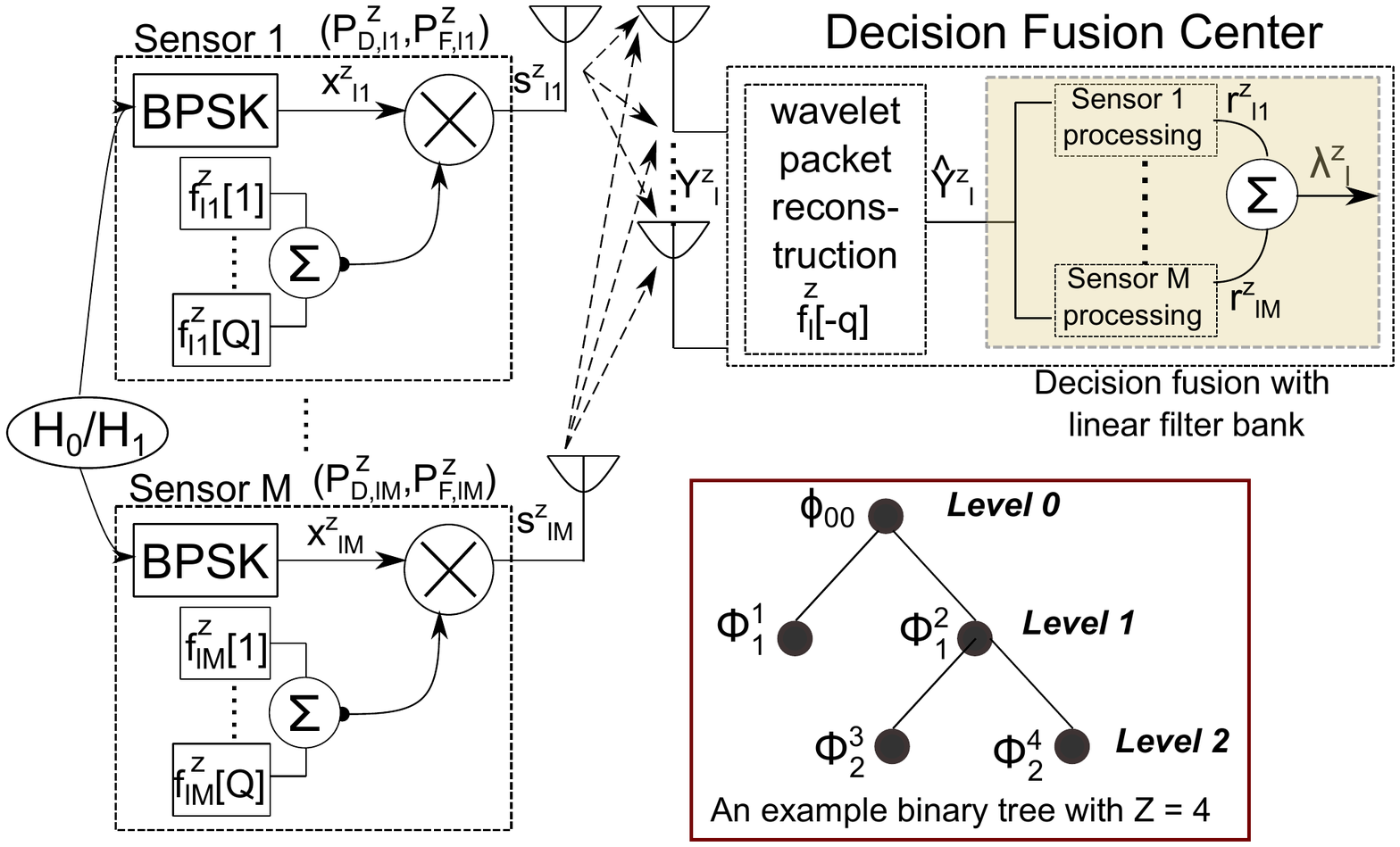}
\end{center}
\vspace*{-85mm}
\caption{Representation of WPDM-aided large-scale industrial sensing network with sensors belonging to the $z$th group sending their observations to a multi-antenna aided DFC.}
\label{fig2a}
\vspace*{-2mm}
\end{figure}

An example scenario with $Z = 4$ sensor groups is presented in the inset of Fig.~\ref{fig2a}. In Fig.~\ref{fig2a}, $\phi_{lz, m}$ is the family of scaling functions derived from the family of wavelets such that these functions form the terminals (leaves) of the tree. These functions can also be referred to as the wavelet packet basis function or simply a wavelet packet. For a given tree structure, the scaling function, $\phi_{lz, m}$ can be constituted using,
\begin{align}\label{eq1}
\mathbf{\Phi}_{l}^z = [\phi_{lz, 1}, \phi_{lz, 2}, \dotso, \phi_{lz, M}] = \sum_{q = 1}^Q \mathbf{f}_{l}^z[q]~\phi_{00}(qT_0)
\end{align}
where $\mathbf{f}_{l}^z[q] = [f_{l1}^z[q], f_{l2}^z[q], \dotso, f_{lM}^z[q]]$ is the equivalent sequence filter built from the combination of the up-sampling and down-sampling finite impulse response (FIR) filters \cite{14} and $\phi_{00}$ is the root of the tree structure. It is worth-mentioning here that $\mathbf{\Phi}_{l}^z$ functions are self and mutually orthogonal at integral multiples of $T_l$ and have a finite duration.

Each of the $M$ sensors in the $z$th sensor group sends its local observation based on a binary hypothesis test, i.e. $\mathcal{H}_{l1}^z$, if the intended target is present, and $\mathcal{H}_{l0}^z$ if it is absent or at a level below the recommended threshold. The individual sensor observations are mapped onto a binary phase shift keying (BPSK)-modulated symbol, such that, $x^z_{lm} \triangleq \{+1, -1\}$. Here the $m$th sensor observation to a $z$th group is mapped onto the $l$th level of the binary tree such that,
\begin{align}\label{eq2}
s^z_{lm} = x^z_{lm} \sum_{q = 1}^Q \sum_{k = 1}^K f^z_{lm}[k]~\phi_{00}((Zq + k)T_0)
\end{align}
where $1 \leq K \leq Q/2$; $Q$ is the (even) total number of filter coefficients, and $K$ is the total number of zeros of the FIR filter, $F^z_{lm}(\zeta)$ at $\zeta = -1$ where $F^z_{lm}(\cdot)$ is the $Z$-transform of $f^z_{lm}[\cdot]$. Therefore, $s^z_{lm} = x^z_{lm} \sigma^z_{lm}[k, q]$ where $\sigma^z_{lm}[k, q] = \sum_{q = 1}^Q \sum_{k = 1}^K f^z_{lm}[k]~\phi_{00}((Zq + k)T_0)$ which makes up the encoded set of sensor decisions on the $l$th level as,
\begin{align}\label{eq3}
\mathbf{s}^z_l \triangleq [s^z_{l1}, s^z_{l2}, \dotso s^z_{lM}]^t = [x^z_{l1}\sigma^z_{l1}[k, q], \dotso, x^z_{lM}\sigma^z_{lM}[k, q]]^t.
\end{align}
Owing to the orthogonality between the wavelet packet basis functions $\mathbf{\Phi}^z_l$, different groups of sensors ($z \triangleq \{1, \dotso, Z\}$) can be time division multiplexed, while arrays can be implemented at the DFC for fusing sensor observations and obtaining a final decision (refer to Fig.~\ref{fig2a}). At the $(l,z)$th terminal (the $z$th group of sensors at the $l$th level), $\mathbf{s}^z_l$ is time-multiplexed with other $(Z - 1)$ group of sensors to be communicated over the wireless channel. The bandwidth of the channel is however assumed to be sufficiently large as not to cause any significant distortion of the waveform of the transmitted sensor observations.

\subsection{Signal and Channel Models}

The signal received at the DFC can be written in the discrete-time matrix form over the $l$th level as,
\begin{align}\label{eq4}
\mathbf{Y}^z_l = \sqrt{\rho_l} \mathbf{G}^z_l \mathbf{s}^z_l + \mathbf{e}^z_l \triangleq [y^z_{l1}, y^z_{l2}, \dotso y^z_{lM}]^t
\end{align}
where $\mathbf{Y}^z_l$ is the received signal vector from the $z$th group of sensors that has been mapped on the $l$th level, $\mathbf{G}^z_l$ is the channel matrix between the $z$th group of sensors and the DFC, $\mathbf{e}^z_l \sim \mathcal{N}_{\mathbb{C}} \big(\mathbf{0}_{N \times T_l},~\Sigma^2_e \mathbf{I}_{N \times T_l}\big)$ is the noise vector and $\mathcal{N}_{\mathbb{C}}(\mu, \varphi)$ denotes complex normal distribution with mean $\mu$ and co-variance $\varphi$. Here $\mathbf{e}^z_l$ accounts for heterogeneous noise encountered in industrial environment including both impulsive and Gaussian noise. Therefore, $\Sigma^2_e = \Sigma_w^2 + \sum_{\gamma = 1}^5 \Sigma_I^2 /\gamma A$ for Middleton Class A noise and $\Sigma^2_e = \Sigma_w^2 + M \Sigma_I^2 /\varrho$ for Bernoulli-Gaussian distributed noise, where $\Sigma_w^2$ is the variance of the Gaussian noise and $\Sigma_I^2$ is the variance of the impulsive noise, $A = \eta \tau/\tau_0$, if the impulsive noise follows the Middleton Class A distribution, $\eta$ is the average number of impulses per second, $\tau_0$ is the interval between two impulses, and $\tau$ is the average duration of each impulse. If the impulsive noise is Bernoulli-Gaussian distributed, it will follow that distribution with a probability of $\varrho$. Each element of the channel matrix $\mathbf{G}^z_l$ is given by $g^z_{n,lm} = \sqrt{\lambda_{lm}}h^z_{n,lm}$ with the geometric attenuation and the shadow fading $\lambda_m$ and fast fading coefficients $h^z_{n,lm}$. Consequently, $\mathbf{G}^z_l = \mathbf{H}^z_l\sqrt{\mathbf{D}_l} \in \mathbb{C}^{N \times M}$ denotes the matrix of the generic channel coefficients, $\mathbf{H}^z_l \in \mathbb{C}^{N \times M}$ denotes the matrix of the fast fading coefficients, and $\mathbf{D}_l$ is a diagonal matrix with $d_{lm,lm} = \lambda_{lm}$.

At the DFC, by applying a multirate filter bank with filters having impulse responses $\mathbf{f}^z_l[-q]$, a process called wavelet packet reconstruction, we can arrive at the linearized equivalent received signal model as,
\begin{align}\label{eq6}
\hat{\mathbf{Y}}^z_l = \sqrt{\rho_l} \hat{\mathbf{G}}^z_l \mathbf{r}^z_l~\hat{\mathbf{\Theta^z_l}}[k, q] + \mathbf{e}^z_l 
\end{align}
where $\hat{\mathbf{G}}^z_l$ is the estimated channel matrix. The DFC estimates the channel state information (CSI), where half of the coherence interval of the channel is used for training to estimate the channel and establish the frequency and timing synchronization. A recommended channel estimator is the minimum mean-squared error (MMSE) estimator. In this paper, we do not separately estimate the channel and calculate the channel estimation error. We just assume that the channel estimation error is included within $\Sigma^2_w$, the variance of the Gaussian distributed noise. In (\ref{eq6}), $\mathbf{r}^z_l = [r^z_{l1}, r^z_{l2}, \dotso, r^z_{lM}]^t$ are the recovered sensor observations in order to obtain the equivalent sequence at the root level $l$, as, $r^z_{l0}[k] = x^z_{lm} \sum_q f^z_{lm}[k - Zq]$. It is possible to recover the original transmitted $m$th sensor observation as, $x^z_{lm} = \sum_k f^z_{lm}[k - Zq] r^z_{l0}[k]$. In (\ref{eq6}), $\hat{\mathbf{\Theta^z_l}}$ is the vector of the auto-correlation functions of the wavelet packets at the $l$th level, expressed as, $\hat{\mathbf{\Theta^z_l}} \triangleq [R_{\phi}(T_0 - qT_0), R_{\phi}(2T_0 - qT_0), \dotso, R_{\phi}(KT_0 - qT_0)]$, where $R_{\phi}(\cdot)$ is the auto-correlation function of $\phi_{00}((Zq + k)T_0)$.

\section{Filters and Fusion}

At the DFC, we employ linear filter-based optimum and sub-optimum decision fusion rules, by recovering decision from each sensor from each group and then fusing them to obtain the final decision for each group of sensors. For example, the received set of vectors at the DFC, $\hat{Y}^1_1, \hat{Y}^2_1, \dotso, \hat{Y}^{Z-1}_{\lceil\log_2Z\rceil}, \hat{Y}^Z_{\lceil\log_2Z\rceil}$ can be analyzed at the DFC to obtain a reliable set of decisions; $\mathcal{H}^1_i$, $\mathcal{H}^2_i$, $\dotso$, $\mathcal{H}^Z_i$ where $i = 1$ or 0 depending on the tested scenario.

\subsection{Design of Fusion Rules}

We start by deriving the optimum fusion rule based on the linear multi-rate filter bank as,
\begin{align}\label{eq8}
&\ln\big[p(\mathbf{r}^z_l|\hat{\mathbf{G}}^z_l\hat{\sigma}^z_{lm}[k, q], \mathcal{H}^z_{i,l})\big] \nonumber\\
&\quad\approx \sum_{m = 1}^M \ln \big[\sum_{x_m}\psi(r^z_{lm}|x^z_{lm})P(x^z_{lm}|\mathcal{H}^z_{i,l})\big]
\end{align}
for parallel access sensor channels \cite{15} where $\psi(\cdot|\cdot)$ is the conditional distribution of individual recovered sensor decision with respect to the transmitted one. In case of multiple access sensor channel, the log-likelihood ratio (LLR) \cite{10} of $\mathbf{r}^z_l$ can be approximated as,
\begin{align}\label{eq9}
\Lambda^z_l \triangleq \ln\Bigg[\frac{p(\mathbf{r}^z_l|\hat{\mathbf{G}}^z_l\hat{\sigma}^z_{lm}[k, q], \mathcal{H}^z_{1,l})}{p(\mathbf{r}^z_l|\hat{\mathbf{G}}^z_l\hat{\sigma}^z_{lm}[k, q], \mathcal{H}^z_{0,l})}\Bigg]
\end{align}
where $\hat{\sigma^z_{lm}[k, q]} = \sum_q \sum_k \hat{f}^z_{lm}[k]R_{\phi}(kT_0 - qT_0 + \Delta)$, $R_{\phi}$ is the auto-correlation function of $\phi_{00}((Zq + k)T_0)$, $P(\cdot)$ and $P(\cdot|\cdot)$ are the individual and conditional probability mass functions respectively and $p(\cdot)$ and $p(\cdot|\cdot)$ are the individual and conditional probability density functions respectively, $\hat{f}^z_{lm}[k]$ is the set of modified filter coefficients and $\Delta$ is the random variable that represents the timing discrepancy between sensor transmissions and reception at the DFC in absence of any synchronization between them. Since $\Delta$ varies slowly over a group of symbols, a constant timing error is visible. Equation (\ref{eq9}) can then be expressed as,
\begin{align}\label{eq10}
&\Lambda^z_l \approx \sum_{m = 1}^M \ln\Bigg[\frac{\psi(r^z_{lm}|x^z_{lm} = 1)P^z_{D,lm} + \psi(r^z_{lm}|x^z_{lm} = -1)(1 - P^z_{D,lm})}{\psi(r^z_{lm}|x^z_{lm} = 1)P^z_{F,lm} + \psi(r^z_{lm}|x^z_{lm} = -1)(1 - P^z_{F,lm})}\Bigg]
\end{align}
where $P^z_{D,lm}$ and $P^z_{F,lm}$ are the probabilities of detection and false alarm of the $m$th sensor on the $l$th layer, with the $m$th sensor belonging to the $z$th group of sensors.

Using linear filter bank processing, we can therefore formulate a set of sub-optimum rules such that,
\begin{align}\label{eq11}
{\mathbf{r}}^z_l = (\mathbf{A}^z_l)^{\dagger} \hat{\mathbf{Y}}^z_l
\end{align}
where $\mathbf{A}^z_l$ can be realized using either Matched Filter (MF) or a zero-forcing (ZF) detector and $(\cdot)^{\dagger}$ denotes the conjugate transpose. In essence,
\begin{align}\label{eq12}
{\mathbf{r}}^z_{l, \text{MF}} &\triangleq  (\hat{\mathbf{G}}^z_l\hat{\sigma}^z_{lm}[k, q])^{\dagger} \hat{\mathbf{Y}}^z_l \nonumber\\
{\mathbf{r}}^z_{l, \text{ZF}} &\triangleq  (\hat{\mathbf{G}}^z_l\hat{\sigma}^z_{lm}[k, q])^{\dagger} \big((\mathbf{\mathcal{D}}^z_l)^{-1}\big)^{\dagger} \hat{\mathbf{Y}}^z_l
\end{align}
where $\mathbf{\mathcal{D}}^z_l = \frac{1}{N}(\hat{\mathbf{G}}^z_l\hat{\sigma}^z_{lm}[k, q])^{\dagger}(\hat{\mathbf{G}}^z_l\hat{\sigma}^z_{lm}[k, q])$ is a diagonal matrix for $N >> M$.

\subsection{Performance Measures}

If we assume that $P(\hat{\mathbf{Y}}^z_l|\hat{\mathbf{G}}^z_l\hat{\sigma}^z_{lm}[k, q], \mathcal{H}^z_{i,l})$ is Gaussian mixture distributed, then we can express $\mathbf{r}^z_l|\hat{\mathbf{G}}^z_l\hat{\sigma}^z_{lm}[k, q], \mathcal{H}^z_{i,l}$ as Gaussian mixture distributed such that, $\mathbf{r}^z_l|\hat{\mathbf{G}}^z_l\hat{\sigma}^z_{lm}[k, q], \mathbf{x}^z_l \sim \mathcal{N}\big(\mathbb{E}\{\mathbf{r}^z_l|\hat{\mathbf{G}}^z_l\hat{\sigma}^z_{lm}[k, q], \mathbf{x}^z_l\}, \mathbb{V}\{\mathbf{r}^z_l|\hat{\mathbf{G}}^z_l\hat{\sigma}^z_{lm}[k, q], \mathbf{x}^z_l\}\big)$ where $\mathcal{N}(\mathbb{E}\{\cdot\}, \mathbb{V}\{\cdot\})$ represents normal distribution with mean $\mathbb{E}\{\cdot\}$ and covariance $\mathbb{V}\{\cdot\}$. Therefore,
\begin{align}\label{eq13}
{\mathbf{r}}^z_{l, \text{MF}}|\hat{\mathbf{G}}^z_l\hat{\sigma}^z_{lm}[k, q], \mathbf{x}^z_l &\sim \mathcal{N}\big(N\mathbf{\mathcal{D}}^z_l\sqrt{\rho_l}\mathbf{x}^z_l, \Sigma^2_e||N\mathbf{\mathcal{D}}^z_l||\big) \nonumber\\
{\mathbf{r}}^z_{l, \text{ZF}}|\hat{\mathbf{G}}^z_l\hat{\sigma}^z_{lm}[k, q], \mathbf{x}^z_l &\sim \mathcal{N}\big(N\sqrt{\rho_l}\mathbf{x}^z_l, \Sigma^2_e||N(\mathbf{\mathcal{D}}^z_l)^{-1}||\big)
\end{align}
where $\mathbf{x}^z_l = [x^z_{l1}, x^z_{l2}, \dotso, x^z_{lM}]$ is the transmit vector of $M$ sensor observations multiplexed on the $l$th level of the wavelet packet binary tree. Now since, $p(\mathbf{r}^z_l|\hat{\mathbf{G}}^z_l\hat{\sigma}^z_{lm}[k, q], \mathbf{x}^z_l) \approx \prod_{m = 1}^M \psi(r^z_{lm}|x^z_{lm})$, we can express (\ref{eq13}) in terms of individual sensor observations assuming they are independent of each other as,
\begin{align}\label{eq14}
\psi(r^z_{lm, MF}|x^z_{lm}) &\triangleq \mathcal{N}\big(Nd^z_{lm}\sqrt{\rho_l}x^z_{lm}, \Sigma^2_e Nd^z_{lm}\big) \nonumber\\
\psi(r^z_{lm, ZF}|x^z_{lm}) &\triangleq \mathcal{N}\big(N\sqrt{\rho_l}x^z_{lm}, N\Sigma^2_e/d^z_{lm}\big)
\end{align}
where $d^z_{lm}$ is the $m$th element of $\mathbf{\mathcal{D}}^z_l$. From (\ref{eq14}), it will be possible to estimate the transmit observations which are BPSK modulated. 

The probability of error in estimating the transmitted sensor observations can be calculated by finding the probability of error in the final obtained decision on whether a target or phenomenon is present ($P^z_{F1,l}$) or absent ($P^z_{F0,l}$) at the $l$th level for the $z$th sensor group. Therefore, 
\begin{align}\label{eq15}
P^z_{F0,l} = \lim_{N \to \infty} \mathcal{Q}\Bigg(\frac{{\mathbf{r}}^z_{l} - \sqrt{N\mathbf{\mathcal{D}}^z_l\rho_l}\mathbf{x}^z_l/\Sigma_e}{\sqrt{1/2(M(1 - P^z_{F,l}) + \Sigma^2_e)}}\Bigg)
\end{align}
where $\mathcal{Q}(\cdot)$ is used to denote the complementary cumulative distribution function (CCDF) and $P^z_{F,l}$ is the probability of wrongly detecting a `0' or absence of a phenomenon or a target.

\section{Performance Analysis}

Before simulating the performance of our designed WPT aided industrial WSN, we need to decide on the design criteria of three components; noise model for $\mathbf{e}^z_l$, the filter coefficients $f^z_{lm}[q]$ and the scaling functions for the wavelets $\phi_{lz, m}$.

\subsubsection{Noise Model}

The combination of Gaussian and impulsive noise follows the pdf, $p(\mathbf{e}) = (1 - \wp)\mathcal{N}(0, \Sigma_g^2) + \wp\mathcal{N}(0, \sqrt{\kappa}\Sigma_I^2/A)$, where $\wp$ denotes the probability of having an impulsive noise with zero mean and variance $\sqrt{\kappa}\Sigma_I^2/A$, while the probability of having a Gaussian distributed background noise is $(1 - \wp)$ with zero mean and variance $\Sigma_g^2$. We also define a factor $\Gamma = \Sigma_g^2/\Sigma_I^2 = 0.25$ implying that the impulsive noise is 25 times stronger than the background Gaussian noise. The variance of the impulse noise is given by $\sqrt{\kappa}\Sigma_I^2/A$ where $\kappa$ is the Poisson distributed sequence whose pdf is characterized by the impulse index $A$. For our case, we choose $A = 0.1$. For Bernoulli-Gaussian distributed noise, the combined noise pdf can be modified to $p(\mathbf{e}) = (1 - \wp)\mathcal{N}(0, \Sigma_g^2) + \wp\mathcal{B}(\varrho, \varsigma\varrho(1 - \varrho))$ where $\mathcal{B}$ represents Bernoulli distribution with mean $\varrho$, variance $\varrho(1 - \varrho)$ and $\varsigma$ is the frequency of occurrence of the impulse noise. For our case, we have $\varrho =0.3$.

\begin{figure}[t]
\begin{center}
 \includegraphics[width=0.6\linewidth]{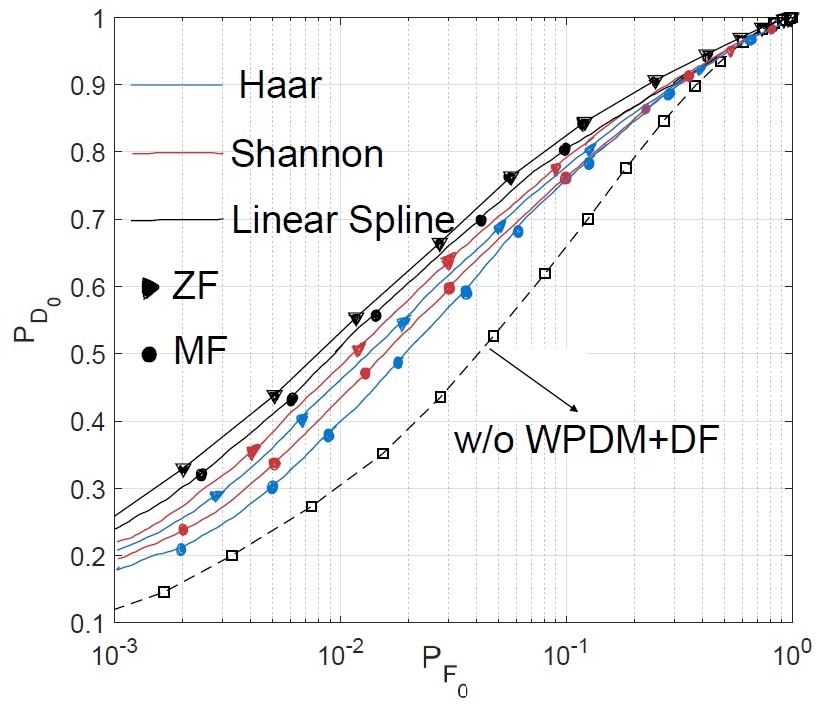}
\end{center}
\vspace*{-5mm}
\caption{Comparative ROC ($P_{D_0}$ v/s $P_{F_0}$) of WPDM and DF aided industrial sensing network with different wavelet scaling functions and fusion rules in presence of Middleton Class A impulse noise with parameters $Z = 4, M = 8, N = 64, \wp = 0.3$ at a fixed SNR of 10dB.}
\label{fig1}
\end{figure}

\begin{figure}[t]
\vspace*{-2mm}
\begin{center}
 \includegraphics[width=0.6\linewidth]{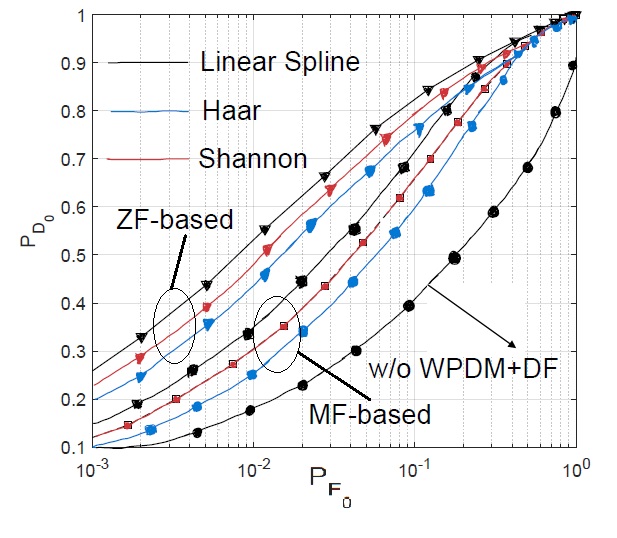}
\end{center}
\vspace*{-5mm}
\caption{Comparative ROC ($P_{D_0}$ v/s $P_{F_0}$) of WPDM and DF aided industrial sensing network with different wavelet scaling functions and fusion rules in presence of Middleton Class A impulse noise with parameters $Z = 4, M = 8, N = 64, \wp = 0.5$ at a fixed SNR of 10dB.}
\label{fig2}
\end{figure}

\subsubsection{Filter Coefficients}

For our synthetic system, we choose an FIR filter of length $Q = 14$ and $K = 2$. The filter bandwidth $B$ is chosen to be equal to $B = \sqrt{2}$ such that $K_0 = 2K - 1 - \lceil 2 \log_2 B \rceil \geq 1$ where $K_0$ is the number of continuous derivatives of the auto-correlation functions, $R_{\phi}(\cdot)$. Using Nyquist sampling rate, the sampling frequency $f_s$, can be calculated as $B = f_s/2$. Using these information, we calculated the filter coefficients, $h[q] = \sinc (q - D)$ and $g[q] = (- 1)^q h[2k + 1 - q]$ where $D$ is the total delay of the FIR filters given by $D = (Q - 1)/4B$, $h[q]$ and $g[q]$ are the filter coefficients for the two branches of the binary tree structure at each level of the tree (refer to Fig.~\ref{fig2a}). The filter coefficients $h[q]$ and $g[q]$ for $q = 0, 1, \dotso, 13$ satisfy the orthonormality and regularity constraints. The equivalent filter coefficients $f^z_{lm}[q]$ can then be built from $h^z_{lm}[q]$ and $g^z_{lm}[q]$, from which the filter sequence $\mathbf{f}^z_l[q]$ can be built for $M$ sensors belonging to the $z$th group of sensors. 

\subsubsection{Scaling Functions}

Three different scaling functions are constructed using standard wavelets from \cite{13}. Performance of our designed transmission system is simulated using the three scaling functions separately, comparing them to find out the set of wavelet scaling functions that offers the best fusion performance under different conditions. The wavelet scaling functions selected are,
\begin{align}\label{eq16}
&\text{Haar :}~~~~\phi_{00}(x) = 1~~~0 \leq x < 1 \nonumber\\
&\text{Shannon :}~~~~\phi_{00}(x) = \sinc (x) \nonumber\\
&\text{Piecewise linear spline :}~~~~\phi_{00}(x) = 1 - |x| 
\end{align}
where $\phi_{00}(x)$ are the root scaling functions which are fed back to (\ref{eq1}) to generate the family of scaling functions to obtain,
\begin{align}\label{eq17}
&\text{Haar :}~\mathbf{\Phi}^z_l = \sum_{q = 1}^Q \mathbf{f}^z_l[q];~\text{Shannon :}~\mathbf{\Phi}^z_l = \sum_{q = 1}^Q \mathbf{f}^z_l[q]~\sinc(qT_0) \nonumber\\
&\text{Piecewise linear spline :}~~~~\mathbf{\Phi}^z_l = \sum_{q = 1}^Q \mathbf{f}^z_l[q](1 - |qT_0|)
\end{align}
Using the scaling functions from (\ref{eq17}), we simulate the performance of a WPT aided industrial WSN. Each of the $Z$ sensor groups are randomly deployed and uniformly distributed in a circular annulus around the DFC with radii $\varphi_{\text{max}} = 1000$~m and $\varphi_{\text{min}} = 100$~m. Within each sensor cluster, $M$ sensors are randomly deployed within a rectangular area (size of 50m $\times$ 100m) following a half-normal distribution \cite{12}. The channel coefficients are log-normal shadowed and Rayleigh faded such that, $h^z_{n,lm} \sim \mathcal{N}_{\mathbb{C}}(0, \text{diag}(B^z_{lm}))$ where $\lambda^z_{lm} = \upsilon_{lm}\big(\frac{\varphi_{\text{min}}}{\varphi_{lm}}\big)^{\eta}$, $B^z_{lm} = \big(\beta^z_{lm}(0), \dotso, \beta^z_{lm}(qT_0)\big)^t$, $10\log_{10}(\upsilon_{lm}) \sim \mathcal{N}\big(\mu_{\lambda}dB, \sigma^2_{\lambda}dB\big)$, $\eta$ is the pathloss exponent, $\varphi_{lm}$ is the distance of the mth sensor from the DFC at the $l$th level of the binary wavelet packet tree, $\rho_l = 1/\sqrt{N}$ and $\{P^z_{D,l}, P^z_{F,l}\} = \{0.5, 0.05\}$.

\begin{figure}[t]
\begin{center}
 \includegraphics[width=0.6\linewidth]{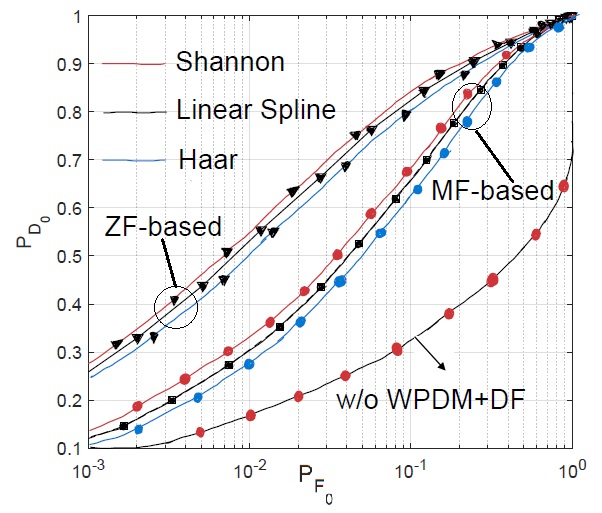}
\end{center}
\vspace*{-5mm}
\caption{Comparative ROC ($P_{D_0}$ v/s $P_{F_0}$) of WPDM and DF aided industrial sensing network with different wavelet scaling functions and fusion rules in presence of Middleton Class A impulse noise with parameters $Z = 4, M = 8, N = 64, \wp = 0.7$ at a fixed SNR of 10dB.}
\label{fig3}
\end{figure}

\begin{figure*}[t]
\begin{minipage}[b]{0.3\linewidth}
\centering
\includegraphics[width=0.99\textwidth]{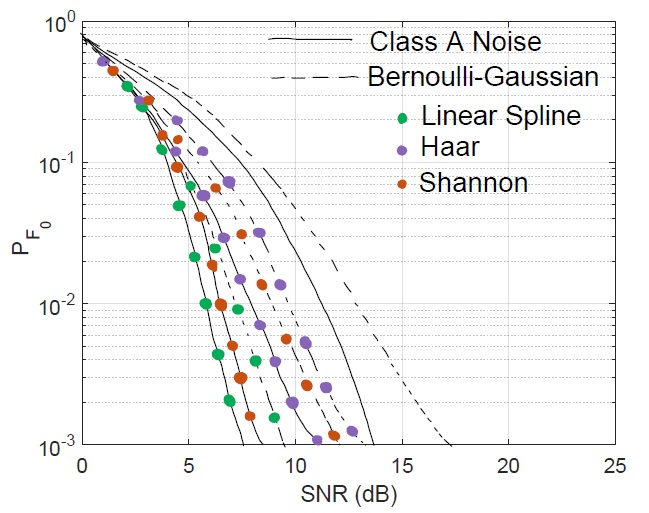}
\vspace*{-6mm}
\caption{Comparative probability of erroneous/false detection as a function of SNR (dB) for ZF-based fusion applied to a WPDM-aided ($Z = 4, M = 8, N = 64$) industrial sensor network with different wavelet scaling functions when the impulse noise ($\wp = 0.3$) is either Middleton Class A or Bernoulli-Gaussian distributed.}
\label{FIG3}
\end{minipage}
\hspace{0.3cm}
\begin{minipage}[b]{0.3\linewidth}
\centering
\includegraphics[width=0.99\textwidth]{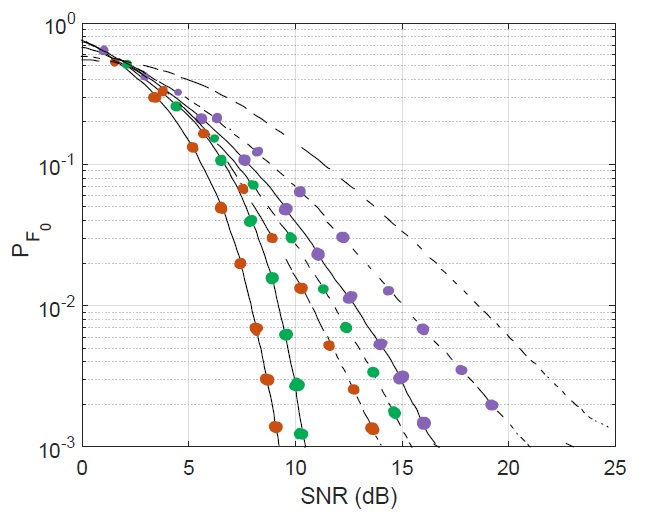}
\vspace*{-6mm}
\caption{Comparative probability of erroneous/false detection as a function of SNR (dB) for ZF-based fusion applied to a WPDM-aided ($Z = 4, M = 8, N = 64$) industrial sensor network with different wavelet scaling functions when the impulse noise ($\wp = 0.5$) is either Middleton Class A or Bernoulli-Gaussian distributed.}
\label{FIG4}
\end{minipage}
\hspace{0.3cm}
\begin{minipage}[b]{0.3\linewidth}
\centering
\includegraphics[width=0.99\textwidth]{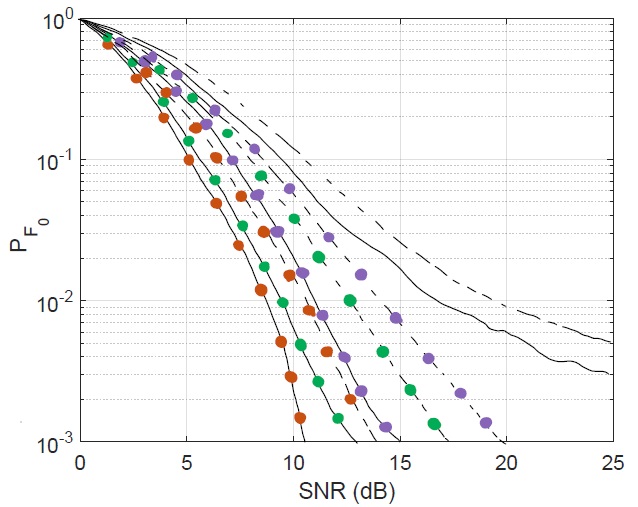}
\vspace*{-6mm}
\caption{{Comparative probability of erroneous/false detection as a function of SNR (dB) for ZF-based fusion applied to a WPDM-aided ($Z = 4, M = 8, N = 64$) industrial sensor network with different wavelet scaling functions when the impulse noise ($\wp = 0.7$) is either Middleton Class A or Bernoulli-Gaussian distributed.}}
\label{FIG4a}
\end{minipage}
\end{figure*}

We generate two sets of figures, (Fig.~\ref{fig1}, Fig.~\ref{fig2}, Fig.~\ref{fig3}) demonstrating receiver operating characteristics (ROC) and (Fig.~\ref{FIG3}, Fig.~\ref{FIG4}, Fig.~\ref{FIG4a}) depicting the probability of false detection against signal-to-noise ratio (SNR) in dB. Both the set of figures are generated for the configurations, $Z = 4, M = 8, N = 64$, thereby representing a virtual large-scale distributed MIMO set-up. In this case, we are considering 4 groups of sensors, each group containing 8 sensors, while the DFC is equipped with 64 antennas. Therefore, the number of levels of the binary tree, $L = \log_2Z = \log_24 = 2$. Also the $T_0 = 10^{-3}$ for the interval between BPSK modulated symbols that are transmitted as the sensor observations; hence $T_l = \{2^1T_0, 2^2T_0\} = \{2 \times 10^{-3}, 4 \times 10^{-3}\}$. Also the pathloss exponent for the MAC channel is 2, with $(\mu_{\lambda}, \sigma_{\lambda}) = (4, 2)$ dB representing indoor smart industrial environment \cite{16}.

All the figures represent the MAC scenario suffering from different combinations of Gaussian and impulsive noise by varying $\wp = (0.3, 0.5, 0.7)$. The performance of the system with $Z = 4, M = 8, N = 64$ where the sensor observations are sent over the MAC without being encoded into wavelet packets with maximal ratio combining (MRC) for fusing the observations at the DFC, is plotted as the benchmark in Fig.~\ref{fig1}, Fig.~\ref{fig2}, Fig.~\ref{fig3}. The benchmark curves are labelled as `w/o WPDM+DF'. For ROCs in Fig.~\ref{fig1}, Fig.~\ref{fig2} and Fig.~\ref{fig3}, three sets of scaling functions are compared, Haar, Shannon and Piecewise linear spline when two different fusion rules are used, MF and ZF. In all cases, ZF offers better fusion performance as compared to MF. The SNR is fixed at 10 dB. Fig.~\ref{fig1}, $\wp = 0.3$ depicts a condition where the impulse noise is low with higher background Gaussian noise. In this case, all the three wavelet functions perform very close to each other and in such a condition, the Haar scaling function can be preferred over the others owing to its low computational complexity. In Fig.~\ref{fig2}, $\wp = 0.5$, presents a condition with equal amount of Gaussian and impulse noise. Piecewise linear spline function offers the best performance. The reason can be attributed to the fact that spline wavelets are formed by linear combination of B-splines \cite{13} and inherit the properties of the basis functions that form the wavelets. Therefore, they are suitable for isolating noise from the original signal depending on the application scenario. In Fig.~\ref{fig3}, impulse noise is higher than the Gaussian noise ($\wp = 0.7$). In this case, the Shannon wavelet performs best, as with complex-valued wavelets based on $\sinc$, it is highly successful in denoising heterogeneous sensor observation signals. 

For the plots of probability of erroneous/false detection, we compare performances of three wavelet scaling functions with the case without WPDM. For the DFC side, we just use ZF-based decision fusion. WPDM is capable  of denoising the sensor signals and offers improvement in performance by more than 10 times in the worst-case high impulse noise scenario ($\wp = 0.7$, Fig.~\ref{FIG4a}) for a particular value of SNR. Transmission of sensor observations aided by WPDM offers a large improvement in performance in a noisy environment. This is because the transmitted waveforms overlap in time, thereby dispersing the impulse noise energy over several symbols at each binary tree terminal. A moderate noise burst that is capable of resulting in erroneous detection is distributed over several waveforms without causing any false detection.

\section{Conclusion}

In order to design a robust large-scale sensing network for the emerging smart Industry/Industry 4.0 environments, we conceive the novel design of combining WPDM of sensor decisions at the transmit side with DF of heterogeneous sensor decisions on the receive side. Evaluating performance of our proposed system in presence of Gaussian and impulsive noise, Rayleigh block fading and indoor shadowing, we demonstrated that WPDM-aided industrial WSN outperforms conventional large-scale multi-antenna WSN arrangements substantially, in presence of impulsive noise particularly. However, in this paper we assume that each sensor group, $z$, consists of an equal number of sensors, $M$. In future, we will relax this assumption to include different number of sensors belonging to each sensor group. We also plan to conduct a detailed study of the wavelet scaling functions and how to optimize their design depending on the environment at hand. 

\bibliographystyle{IEEEtran}

\end{document}